\documentclass[twocolumn]{article}
 
\usepackage{graphics}

\newdimen\headerboxheight
\newdimen\betweenumberspace          
\newdimen\aftertext                  
\newdimen\headlineindent             
\renewcommand{\dj}{\hbox{d\hskip-1.1ex{\raise0.640ex\hbox{--}}\skip 0.70ex}}
\newcommand{\calQ}{\mathcal{Q}}

\newcommand{\calL}{\mathcal{L}}
\newcommand{\bra}[1]{\langle #1 |}
\newcommand{\ket}[1]{| #1 \rangle}

\if@mathematic
   \def\vec#1{\ensuremath{\mathbf{#1}}}
\else
   \def\vec#1{\ensuremath{\mathchoice{\mbox{\boldmath$\displaystyle#1$}}
                              {\mbox{\boldmath$\textstyle#1$}}
                              {\mbox{\boldmath$\scriptstyle#1$}}
                              {\mbox{\boldmath$\scriptscriptstyle#1$}}}}
\fi

\newcommand{\lsim}[1]{
\setlength{\unitlength}{12pt}
\begin{picture}(1.4,1.)
\put(.7,-0.3){\makebox(0.0,1.)[t]{$<$}}
\put(.7,-0.3){\makebox(0.0,1.)[b]{$\sim$}}
\end{picture}#1}

\begin{document}
 
\twocolumn[

\vspace{3em}

\textbf{\textsf{\LARGE On the instanton-induced portion of the nucleon
strangeness}}

\vspace{1em}

D. Klabu\v{c}ar$^{1a}$ , K. Kumeri\v{c}ki$^{1b}$, 
B. Meli\'{c}$^{2c}$ and I. Picek$^{1d}$

\vspace{1em}
{\small
$^1$ Department of Physics, Faculty of Science, 
      University of Zagreb, POB 162, HR-10001 Zagreb, Croatia \\
$^2$ Theoretical Physics Division, Ru\dj er Bo\v{s}kovi\'{c} Institute,
 Bijeni\v{c}ka cesta 54, HR-10001 Zagreb, Croatia 
}
\vspace{3em}
]
\renewcommand{\thefootnote}{\alph{footnote}}
\stepcounter{footnote}
\footnotetext{klabucar@phy.hr}
\stepcounter{footnote}
\footnotetext{kkumer@phy.hr}
\stepcounter{footnote}
\footnotetext{melic@thphys.irb.hr}
\stepcounter{footnote}
\footnotetext{picek@phy.hr}
\renewcommand{\thefootnote}{\arabic{footnote}}
\setcounter{footnote}{0}

\subsection*{Abstract}

We calculate the instanton contribution to the proton strangeness in 
the MIT bag enriched by the presence of a dilute instanton liquid.
The evaluation is based on expressing the nucleon matrix
elements of bilinear strange quark operators in terms of a model
valence nucleon state and interactions producing quark-antiquark
fluctuations on top of that valence state. Our method combines the
usage of the evolution operator containing a strangeness source, 
and the Feynman-Hellmann theorem. { It enables one to evaluate
the strangeness in different Lorentz channels in, essentially, the
same way. Only the scalar channel is found to be affected by the 
 interaction induced by the random instanton liquid.}

\vskip0.5cm\hrule\vskip3ptplus12pt\null

\section{Introduction}
\label{sec1}

\begin{figure}
\centerline{\resizebox{0.45\textwidth}{!}{\includegraphics{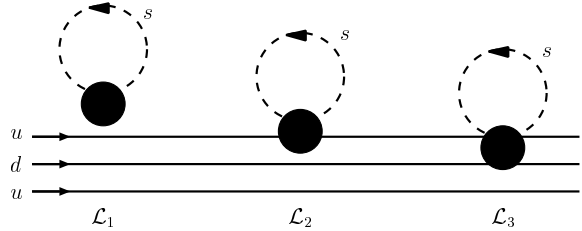}}}
\caption{Instanton-induced local strangeness represented by the
effective one-, two- and three-body operators. Non-strange quarks are
denoted by solid lines, and strange ones by dashed lines.\label{figb}}
\end{figure}

Despite the accumulated evidence for the nucleon stran\-ge\-ness,
there has been as yet no balanced understanding of its various 
appearances. By a particular nucleon strangeness we understand
the value of the nucleon matrix element $\bra{N} \mathcal{O}_{s}
(\Gamma) \ket{N}$, where the bilinear $\mathcal{O}_{s}(\Gamma)
=\bar{s}\Gamma s$ might represent the scalar, pseudoscalar, vector,
axial vector and tensor strange current densities ($\Gamma=1,
\gamma_{5}, \gamma_{\mu}$, $\gamma_{\mu}\gamma_{5}, \sigma_{\mu\nu}$).
Thus, any interaction $\calL_{I}$ that induces $s\bar{s}$ pairs in
the nucleon state $\ket{N}$ potentially leads to various types of
nucleon strangeness. The imaginable interactions $\calL_{I}$,
which are related to QCD-vacuum fluctuations, might prefer
some of the strangeness channels. In particular, there is a conjecture
\cite{Zh97} that a non-trivial QCD-vacuum structure selects the
pseudoscalar and scalar channels, which experience the axial
and trace anomaly, respectively. In the present paper we focus
on QCD-vacuum fluctuations as given by
the instanton-liquid model \cite{ShVZ80,DiP84,NoVZ89}, \emph{i. e.}
we take $\calL_{I}\rightarrow\calL_{\rm inst}$. 
Such an interaction generates
an $s$-quark loop (schematically shown in Fig. \ref{figb}) 
to which an
external probe can couple. It is important that this interaction
can be treated perturbatively and enables us to compare its
relative contributions to different strangeness channels.
The relatively complicated interaction $\calL_{I}=\calL_{\rm inst}$ 
[given by Eqs. (\ref{eq:L1})--(\ref{eq:L3}) below] 
is conveniently split into three pieces:
\begin{equation}
\calL_{I}=\calL_{1}+\calL_{2}+\calL_{3} \;,
\label{Linst}
\end{equation}
where the parts illustrated in Fig. \ref{figb} 
refer to the one-, two- and three-body
operators. These operators  change the known valence (model)
state $\ket{N_{0}}$ to the state $\ket{N}$ containing the
$s\bar{s}$ pairs. Then, we provide an expression (Eq.
(\ref{formula})) suitable for
computing the strange matrix element of the full nucleon state,
$\bra{N}:\bar{s}\Gamma s :\ket{N}$.

\begin{figure}
\centerline{\resizebox{0.40\textwidth}{!}{\includegraphics{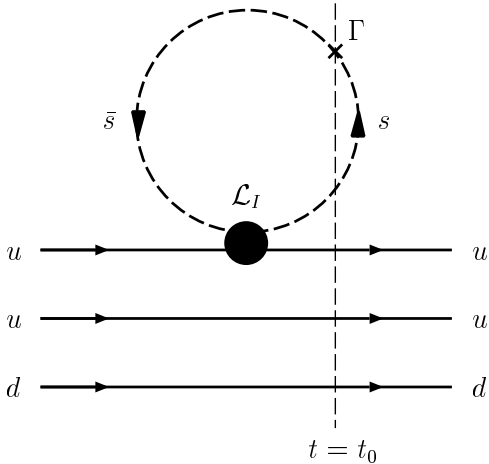}}}
\caption{Non-vanishing nucleon strangeness due to 
a response of the valence nucleon state 
to a strangeness source at $\Gamma$ (denoted by $\times$), {\it i.e.} to 
a probe coupled to strange quarks through $\Gamma$.
More precisely, this graph is that part of the nucleon response which
arises only through one interaction ${\cal L}_I$.\label{fig1}}
\end{figure}

The current evidence for the strangeness content of the proton comes
from the external probe both at low- and at high-momentum transfers.
The analysis of the term $\sigma_{\pi N}$ in low-energy $\pi N$ scattering 
reveals  comparable light and strange quark nucleon matrix
elements \cite{Zh97} ($N$ means the proton throughout this paper):
\begin{eqnarray}
\bra{N}\bar{u}u\ket{N}&\simeq & 4.8\;, \nonumber \\
\bra{N}\bar{d}d\ket{N}&\simeq & 4.1\;, \nonumber \\
\bra{N}\bar{s}s\ket{N}&\simeq & 2.8\;,
\end{eqnarray}
{\it i.e.}, the unexpectedly large \emph{scalar strangeness}.
A posteriori, it is found to be in accordance with QCD-vacuum
characteristics \cite{Zh97}, as represented, for example, by
the (na\-i\-ve) bag-model relation \cite{DoN86}
\begin{equation}
\bra{N}\bar{s}s\ket{N} = - \bra{0}\bar{s}s\ket{0} V
\label{naive}
\end{equation}
or the QCD sum-rules result \cite{KhKZ87}
\begin{equation}
\bra{N}\bar{s}s\ket{N}\simeq  2.4 \;.
\end{equation}
The other piece of evidence for the strangeness content of the proton comes
from the polarized lepton-nucleon scattering
at relatively high-momentum transfer, 
{ higher than the scales pertinent for our considerations.} 
The analysis \cite{ElK95} of new data
supports the original EMC findings \cite{As88,As89} --- it
reveals a non-vanishing fraction $\Delta s=-0.11\pm 0.06$ of the proton spin 
$S_{\mu}$ carried by the $s$ quark. This is not negligible in comparison 
with $\Delta u=0.82\pm 0.06$ and $\Delta d=-0.44\pm 0.06$. $\Delta s$ 
is related to the \emph{axial strangeness} of the proton defined as
$\bra{N}\bar{s}\gamma_{\mu}\gamma_{5}s\ket{N}=\Delta s\, S_{\mu}$.

The \emph{vector stran\-ge\-ness}, described by the Dirac and the Pauli 
form factors as
\begin{equation}
\bra{N}\bar{s}\gamma_{\mu}s\ket{N}=\bar{u}_{N}(p')\left[
F_{1}^{s}(q^2)\gamma_{\mu} + F_{2}^{s}(q^2)
\frac{i\sigma_{\mu\nu} q^{\nu}} {2 M_{N}}\right]u_{N}(p),
\label{f1f2}
\end{equation}
can be related to the analogous flavour singlet (0) and the hypercharge 
(8) form factors for $\bra{N} V_{\mu}^{(0,8)} \ket{N}$ through
\begin{eqnarray}
\bar{s}\gamma_{\mu}s&=&V^{(0)}_{\mu}-2 V^{(8)}_{\mu} \;, \nonumber \\
V^{(0)}_{\mu}&=&\frac{1}{3}\bar{q}\gamma_{\mu}q\;,\quad
V^{(8)}_{\mu}=\frac{1}{\sqrt{3}}
\bar{q}\gamma_{\mu}\frac{\lambda^{8}}{2}q \;.
\end{eqnarray}
Although $F_{1}^{s}(0)=0$ as the net nucleon strangeness, its
momentum dependence determines the strangeness radius
\begin{equation}
r_{s}^{2}=6\frac{d}{dq^2}F_{1}^{s}(q^2)\Bigg|_{q^2=0} \;,
\end{equation}
while the strange magnetic moment is given by
\begin{equation}
\mu_{s}=F_{2}^{s}(0) \;.
\end{equation}
Note the relation $F_{2}^{s}=F_{2}^{(0)}-\frac{2}{\sqrt{3}}
F_{2}^{(8)}$, where the last term is constrained by
$\frac{2}{\sqrt{3}}F_{2}^{(8)}=\kappa_{p}+\kappa_{n}=-0.12$.
Knowledge of the $F_{2}^{(0)}$ flavour singlet term would also
enable one to determine the baryomagnetic moment
\begin{equation}
\mu_{p}^{(0)}=F_{1}^{(0)}(0)+F_{2}^{(0)}(0)=(1+F_{2}^{(0)})
\; \textrm{n. m.}
\label{baryomagnetic}
\end{equation}

There are many various model-dependent calculations  
\cite{Ja89,KlP89,KlP90,PaSW91,MuB94,FoNJC94,HoPM97,HoP93,GeI97,Le96,Le95,HaMD96,Ch96}
of the quantities listed above.
{ By deriving Eqs. (\ref{formula})--(\ref{eq:contractions}) in the 
next section, we provide a framework which is rather general in 
that it can be applied to different quark models and flavor-mixing
interactions ${\cal L}_I$. We illustrate its usage on the example 
of the MIT bag model and instanton-induced interaction. At least 
in principle, this framework also
treats all Lorentz channels on an equal footing, depending on which 
$\Gamma$ is plugged in Eqs. (\ref{formula})--(\ref{eq:contractions}).
}

\section{Nucleon strangeness induced on top 
of the valence quark state}
\label{sec2}

It is not very surprising in non-perturbative QCD, in the light of
its non-vanishing quark scalar condensates, that some
matrix elements $\langle N |\overline {s}\Gamma s|N\rangle$ 
can be markedly different from zero.
The vacuum expectation value of ${\bar s}s$ is 
actually approximately as large as for non-strange quarks:
$\langle 0|\bar{s} s| 0\rangle \approx \langle 0|\bar{u} u| 0\rangle
= \langle 0|\bar{d} d| 0\rangle$, {\it i.e.} roughly equal to 
or even more negative than $(-200 {\rm MeV})^3$.
The MIT bag mo\-del provides a good illustration how this leads 
to a large $\langle N|\bar{s}s| N\rangle$ \cite{DoN86}. However,
there may also be $s\bar s$-pairs other than those from 
the QCD-vacuum condensate, so that normal-ordered
strange operators can, in principle, also have non-vanishing 
nucleon matrix elements.

{ Since we are interested in the $s\bar{s}$-pairs that may exist in 
addition to those from the (non-\-per\-tur\-ba\-ti\-ve-)va\-cu\-um channel,
it is convenient to define the normal ordering 
with respect to the non-perturbative vacuum $|0\rangle$: 
\begin{equation}
   :\bar{q}\Gamma q: \: = \: \bar{q} \Gamma q - 
          \langle 0 | \bar{q} \Gamma  q | 0 \rangle \;.
\label{defNO}
\end{equation}
Ideally, this referent vacuum state $| 0 \rangle$ would be 
the true non-perturbative vacuum of QCD, 
but since in this paper we are concerned 
with quark models imitating QCD, Eq. (\ref{defNO}) 
will in practice mean that the
normal ordering is taken with respect to a model vacuum state. By this we
mean the ground nucleon state from which the valence quarks are removed 
(for example, the ``empty bag'' in the case of the MIT bag model). 
For the strange quarks, the normal ordering 
with respect to this referent vacuum state $| 0 \rangle$ is
equivalent to the normal ordering 
with respect to the model nucleon ground state $ | N_0 \rangle$
composed of the non-strange valence quarks only. 
Of course, such a definition of normal ordering is then 
necessarily tied to the characteristic hadronic scale of $\sim 1$ GeV, 
at which the non-perturbative QCD effects dominate and at which, and 
below which, quark models provide a reasonable description of the nucleon 
bound state.}
 
Figure \ref{fig1} illustrates how a non-vanishing value not only of 
$\langle N| \bar s \Gamma s |N\rangle$, but also of the normal-ordered 
$\langle N| :\bar s \Gamma s: |N\rangle$ can then come about: 
at the instant $t=t_0$ the composite nucleon is hit by an
external probe ({\it e.g.} a neutrino \cite{Ah87}) with the coupling 
$\Gamma$ to strange quarks. Owing to an interaction capable of 
producing $s\bar s$ fluctuations, the nucleon state $|N\rangle$
at the time slice $t=t_0$ obviously contains not only the valence
quarks $uud$, but also the $s$-quark loop to which the
external probe can also couple. 

Let us {\it schematically} write down the full nucleon (proton) state,
which is also coupled to the strangeness-sen\-si\-ti\-ve probe:
\begin{eqnarray}
 |N\rangle
    & = &  \frac {1}{\cal N} \left(
     \sum_{X=0}^\infty C_{X} |uud \, X \rangle +
     \sum_{X=0}^\infty C_{s\bar s X} |uud \, s\bar {s}\, X\rangle \right)
\nonumber \\
 &\equiv & \frac {1}{\cal N} (|N_{0}\rangle+| \delta N\rangle)
\, . \label{fullproton}
\end{eqnarray}
Here, $X$ (starting from $X=|0\rangle$, the
complicated non-per\-tur\-ba\-ti\-ve QCD vacuum) symbolizes any number of 
various perturbative {\it and} non-per\-tur\-ba\-ti\-ve gluon and 
quark configurations 
including quark-antiquark pairs and, in particular, strange pairs
which escaped detection by this probe.
 These complicated configurations ``dress" quarks ($q=u$, $d$, $s$, ...)
into their effective counterparts --- constituent quarks
${\cal Q}$ = $\mathcal{U}$, $\mathcal{D}$, $\mathcal{S}$, ... .
In terms of the constituent quarks, this part, unperturbed
by the strangeness-sensitive probe, is just the valence part
$| N_0 \rangle = |{\cal U U D} \rangle$ when the nucleon is a proton.
It is obvious in terms of the constituent quarks
that $\langle N_0| :\bar s \Gamma s: |N_0\rangle = 0$.
The {\it one} strange pair detected at $\Gamma$
has been explicitly denoted by $s\bar s$ in the
$|\delta N\rangle$-part of the nucleon state perturbed by the
probe. $|\delta N\rangle$ can be viewed as the response
of $|N_{0}\rangle$ to the weakly coupled strangeness-sensitive probe.
The coefficients $C_X, C_{s\bar s X}$ denote the amplitudes of states
with various admixtures $X$ or ${s\bar s\, X}$.
${\cal N}$ is the normalization.
This response makes possible that the total nucleon
$\Gamma$--strangeness $\langle N| \bar s \Gamma s |N\rangle$
also receives a non-vanishing contribution from the non-vacuum
channel $\langle N| :\bar s \Gamma s: |N\rangle$.

However, the question is how to get the nucleon state  
in sufficiently specific terms in order to have a calculable expression 
for $\langle N |\!:\!\overline {s}\Gamma s\!:\!|N\rangle$.
A viable approach is to resort to a constituent model of hadrons. 
The idea of a constituent model is that all the mess of fluctuations 
$X$ is by some model parameterization lumped into dressing of 
constituent quarks $\cal Q$,
as well as into effective model interactions, or a mean field they feel.
The valence proton state $|N_0\rangle$ would then
be identified with the model ground eigenstate $|\cal UUD\rangle$
built up {\it only} of non-strange effective quarks 
(so that $\langle N_0|\!:\!\overline s \Gamma s\!:\!|N_0\rangle=0$,
even though  possibly $\langle N_0 |\overline{s}\Gamma s|N_0\rangle \neq 0$, 
at least for $\Gamma = 1$, owing to the strange vacuum condensate).
Let us denote all possible higher eigenstates of some model Hamiltonian 
${H}_{0}$ by $|k\rangle$:
\begin{equation}
      {H}_{0}|N_{0}\rangle = E_{N_{0}}|N_{0}\rangle,
\quad 
                   {H}_{0}|k\rangle = E_{k}|k\rangle,
\quad  E_{k} >  E_{N_{0}}.
\label{Spectrum}
\end{equation}
The Hamiltonian ${H}_{0}$ is responsible for the
formation of (model) hadron states composed of definite, fixed
numbers of quarks.
In the simplest case, we can imagine $H_0$ as  consisting
of a sum of one-body quark operators, say
typically of the effective quark kinetic energy operator
and the mean, or self-consistent,  field in which the
dressed valence quarks would move.
For example, $ {H}_{0}$ could be the static bag-model Hamiltonian. 
$|N_{0}\rangle$ would then be the bag-model nucleon in its
ground state, and $|k\rangle$ all higher bag states with a definite
number of constituents. 
In any case, $H_0$ defines the nucleon model --- possibly together with
some other ingredients (such as the confining boundary condition in bag
models).

What $H_0$ cannot do is to produce $s\bar{s}$
fluctuating pairs. To produce such pairs, we have to  
supplement $H_0$ defining the model one starts from, by 
some Hamiltonian ${H}_{I}$ (corresponding to the Lagrangian 
density ${\cal L}_{I}$) which can produce $s\bar{s}$
excitations on top of $|N_{0}\rangle$. This means that ${\cal L}_{I}$,
and thus also ${H}_{I}$,
contains strange quark field operators bilinearly, so that it 
can connect $|N_{0}\rangle$ with  $|\delta N\rangle$ containing 
$s\bar{s}$ pairs. 

{To clarify that introducing ${\cal L}_{I}$ does not lead 
to double-counting, let us repeat that $H_0$ is just a 
model Hamiltonian, the parameters of which should mimic the 
effects of full, true non-perturbative QCD as much as possible. 
For example, if $H_0$ is the
Hamiltonian of the non-relativistic naive constituent quark
model, it must contain the postulated mass parameter of
the constituent quark mass $M_\mathcal{Q} \approx M_{N_0}/3$. The
corresponding quantity in the true theory, the dynamically
generated quark mass, is (in principle) the result of all
possible QCD interactions, so that the interactions related to
$H_I$ can, in real QCD, also contribute to this mass by contributing 
to the $s\bar s$-fluctuations. The
dynamically generated non-strange quark mass must be close to
the model constituent quark mass parameter $M_\mathcal{Q}$ sitting in
$H_0$, and only in such implicit, indirect ways are interactions
like $H_I$ ``present" in $H_0$. However, they are not present
explicitly, and, in fact, $H_0$ cannot produce any $s\bar s$
fluctuations at all. Therefore, if we want to study the $s\bar s$
fluctuations, we must introduce $H_I$ to enrich the model nucleon
with ${\cal S} {\bar{\cal S}}$-fluctuations on top of $|N_0\rangle$.}

In order to obtain the expression for 
$\langle N| :\bar s \Gamma s: |N\rangle$
by utilizing the Feynman-Hellmann theorem \cite{Fe39,He37},
let us define an auxiliary perturbation Hamiltonian 
$ {H}^\prime$ by adding to $ {H}_{I}$ a source term
for the strange operator we want to calculate in the ``full"
nucleon state $|N\rangle$:
\begin{equation} 
     {H}' \equiv 
{H}_{I}+ \lambda \otimes < \overline{s} \Gamma s > ,
\label{Hprime}
\end{equation}
\noindent where $< \overline{s} \Gamma s >$ is the convenient abbreviation
\begin{equation}
<\overline{s} \Gamma s > \equiv 
\int  \overline{s} (x) \Gamma s(x)\, d^{3}x \, . 
\label{sdens}
\end{equation}

\noindent 
The generic form $\lambda \otimes \Gamma$ can mean any of the cases 
$\lambda {\bf 1}_{4}$, $\lambda_{\mu} \gamma^{\mu}$, 
$\lambda _{5 \mu} \gamma^{\mu} \gamma_{5}$, 
$\lambda_{\mu\nu} \sigma^{\mu \nu}$, etc. 

Then we use the auxiliary perturbation Hamiltonian
as the interaction Hamiltonian in the evolution operator 
$U(t_{2},t_{1})$. The perturbation expansion of this operator is
\begin{eqnarray}
{U} (t_{2},t_{1}) & = & 
1 + \sum^{\infty}_{n=1}
{U}^{(n)}(t_{2},t_{1}) \nonumber \\
 & = & \hat{T} \bigg\{ 1+ \sum^{\infty}_{n=1}\frac{i^{n}}{n!}
\Big[\int^{t_2}_{t_{1}}\! :\! L_{\rm{int}}(t)\! : dt \, \Big]^{n} 
\bigg\}  .
\label{EvolOp}
\end{eqnarray}
Here, $\hat{T}$ denotes the time-ordering operator and 
$L_{\rm{int}}(t)= \int {\cal L}_{\rm{int}} ({\bf{x}},t)d^{3}x
= - H_{\rm{int}}(t)$ is the interaction Lagrangian to be replaced
with the form containing the strangeness sources, as in the definition of 
$ {H}^\prime$ (Eq. (\ref{Hprime})):
\begin{eqnarray}
L{(t)}_{\rm{int}} \rightarrow L'(t) & = & 
L_I(t) - \lambda \otimes < \overline{s} \Gamma s (t)>  \nonumber \\
 & = & \int d^{3}x \Big[ {\cal L}_{I}(x)- \lambda \otimes 
\overline{s} (x) \Gamma s (x)\Big].
\label{Lprime}
\end{eqnarray}

The Feynman-Hellmann theorem then enables one to understand 
the nucleon matrix elements of the strange current densities,
$\langle N|:\overline{s} \Gamma s:|N\rangle$, 
as the response (to the strange current source) of the $\langle N_0
(t\rightarrow +\infty)|N_0 (t\rightarrow -\infty)\rangle$
transition amplitude of the model ground state $| N_0 \rangle$.
For example, in the case of the second-order term in Eq. (\ref{EvolOp}),
the substitution (\ref{Lprime}) leads to
\begin{eqnarray}
\lefteqn{{U}^{(2)} (+ \infty, -\infty)\! =\! - \frac{1}{2}\, \hat{T} \int^{+
\infty}_{- \infty}\! dt\! \int^{+\infty}_{-\infty}\! dt' \,
\bigg[\, :\! L_{I}(t)\! :\, :\!  L_{I}(t')\! : }\nonumber \\
&-& \lambda_{\alpha} :<\bar{s} \gamma^{\alpha} s(t)>: \,
:\!  L_{I}(t')\! :
-:\! L_{I}(t)\! : \, \lambda_{\beta}
:<\bar{s} \gamma^{\beta} s (t') >: \nonumber \\
&+& \lambda_{\alpha} \lambda_{\beta} \,
:<\bar{s}\gamma^{\alpha} s(t)>:\, :<\bar{s} \gamma^{\beta}s(t')>:\,\bigg] \;.
\label{U2}
\end{eqnarray}
For definiteness, the above expression for ${U}^{(2)}$ has been written 
for the {\it vector} strange current density.
The first-order contribution to the vector nucleon strangeness 
can then be obtained by considering
\begin{eqnarray}
\frac{\partial}{\partial \lambda_{\mu}}
\langle N_{0} |U^{(2)}
(+ \infty, - \infty)|N_{0}\rangle\Bigg|_{\lambda_{\mu}=0} \, .
\end{eqnarray}

In general, for any matrix $\Gamma$ in the spinor space, the strange 
nucleon matrix element of the full nucleon state $|N\rangle$ is, 
to the two lowest orders (due to the $U^{(2)}$ and $U^{(3)}$ 
terms), given by 
\begin{eqnarray}
\lefteqn{\langle N|:\overline{s} \Gamma s:|N\rangle \, 
= \, i \, 
\int^{+ \infty}_{- \infty}\! dt' \,
\langle N_{0}|{\hat T} :<\overline{s}\Gamma s (t_0)>:} \nonumber \\
& \times & :\! L_{I} (t')\! : |N_{0}\rangle 
 \; - \; \frac{1}{2}
 \int^{+\infty}_{- \infty}\! dt' \,  \int^{+\infty}_{- \infty}\! dt'' 
\nonumber \\
&\times &
\langle N_{0}|{\hat T}:<\overline{s} \Gamma s(t_0) >:\,:\! L_{I}(t')\! :\, 
 :\! L_{I}(t'')\! :|N_{0}\rangle  . \hspace*{5mm}
\label{formula}
\end{eqnarray}

\begin{figure}
\centerline{\resizebox{0.45\textwidth}{!}{\includegraphics{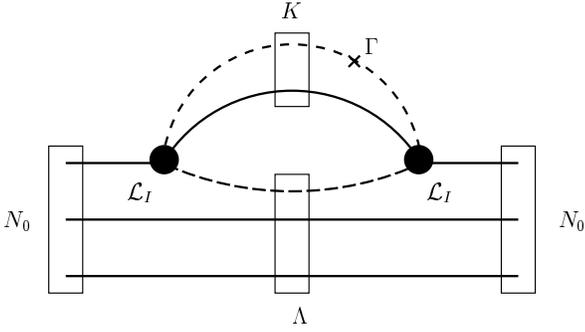}}}
\caption{
A response of the valence nucleon state $|N_0\rangle$
to a strangeness source at $\Gamma$ through two interactions 
${\cal L}_I$. This type of contribution can be associated with 
the kaon-loop contribution to the nucleon strangeness
(a possible $K\Lambda$ intermediate state is therefore indicated).
\label{fig2}}
\end{figure}

Obviously, the non-vanishing contributions to (\ref{formula})
occur only when the strange quark fields are fully contracted. 
For example, the integrand of the first term  in
(\ref{formula}), written in terms of space integrals
over the contracted strange current and Lagrangian densities, is
\begin{eqnarray}
\lefteqn{\int \! d^{3}\! x  \,  d^{3}\! x' 
\langle N_{0}|{\hat T} :\overline{s} (x)
\Gamma s (x): \, :{\cal L}_{I} (x'): |N_{0}\rangle }\hspace*{1cm} \nonumber \\
&=& \int \! d^{3}\! x \, d^{3}\! x' \langle N_{0}| :\overbrace{\overline{s}
(x) \Gamma \underbrace{s(x) {\cal L}_{I}}}
(x'): |N_{0}\rangle \, , \hspace*{1cm}  \label{eq:contractions}
\end{eqnarray}

\noindent where the contractions are indicated by over- and
underbraces, and $t_0 \equiv x_0$ and $t'\equiv {x'}_0$, for consistency
of the notation. So, the first term in (\ref{formula}) corresponds
to Fig. \ref{fig1}, since these contractions, 
or time-ordered pairings, are, of course,
the propagators of strange quarks. In the second term, the two contractions
must connect the strangeness source at $\Gamma$ with two different 
separately normal-ordered interaction Lagrangian densities
which act as ``sinks" for strangeness at two different points of
the valence-quark lines.
In any case, there must be an additional strange-quark contraction between 
these two $:{\cal L}_I:$'s, and this completes the  strange-quark loop.
Fig. \ref{fig2} shows an example of the graphs originating from the second
term of (\ref{formula}), namely the $U^{(3)}$ contribution. 
{ Clearly, in this way, one can generate contributions corresponding
to kaon-baryon loops in the approaches employing hadron degrees of
freedom. Below, we will use a strangeness-generating interaction 
${\cal L}_I$ which is perturbative, so that we do not expect sizable 
contributions to the $s\bar s$-effects from the second order in 
${\cal L}_I$.
In addition to that, there are indications that these contributions
related to strange meson loops, should be rather small even when one
does not restrict oneself to perturbative $s\bar s$-generating 
interactions. Some of these indications come from model-dependent 
calculations, {\it e.g.} in the Nambu and Jona-Lasinio (NJL) model 
\cite{StW94}. Recently, however, Geiger and Isgur presented a 
parameter-free analysis within a rather general framework consistent
with the many empirical constraints (such as OZI rule), which 
shows that a complete set of strong strange meson-baryon loops,
computed in a model consistent with the OZI rule, gives (after
delicate cancellations) only small observable $s\bar s$ effects. 
We therefore do not consider the $s\bar s$-effects from the second 
order in ${\cal L}_I$.
}

\section{Strangeness evaluation with a specified interaction ${\cal L}_{I}$}
\label{sec3}

The evaluation of Eq. (\ref{formula}) is in principle
straightforward once one specifies two things. 
{The first is the overall description of hadronic structure, 
which in practice amounts to choosing the model for the nucleon 
state $|N_0\rangle$ --- for example, choosing some mean-field 
Hamiltonian such as $H_0$ in (\ref{Spectrum}). 
The second is the choice of the interaction (we call it ${\cal L}_{I}$) 
which has the role to generate $q\bar q$ fluctuations on 
top of the valence nucleon state $|N_0\rangle$ (the eigenstate of $H_0$).
Specifying $H_0$ also defines the model single-quark solutions,
and we can use them as an appropriate wave-function basis to
expand the quark fields $q(x)$ ($q=u,d,s$) in terms of creation
(${\cal U}^\dagger_{K}, {\cal D}^\dagger_{K}, {\cal S}^\dagger_{K}$)
and annihilation (${\cal U}_{K}, {\cal D}_{K},{\cal S}_{K}$) 
operators of dressed quarks and antiquarks:
\begin{equation}
q(x)=\sum_K \left( \calQ_K\, q_{K}
( { \vec{x}})e^{-i\omega_K t} +
  \calQ^{c^\dagger}_K\, q_{K}^{c}
( { \vec{x}})e^{ i\omega_K t}
        \right) \;.
\label{expandq}
\end{equation}
Here, $q_{K}(\vec{r})$ denotes a model wave function of a quark of 
flavour $q$, where
$K$ stands for the set of quantum numbers labelling a model quark state.
(For example, in the next section we will choose to employ the MIT bag model.
Then, $q_{K}(\vec{r})$ will be the solution for the quark in the $K$-th
mode of the MIT bag.)} 

The field contractions in (\ref{formula}) lead to the sums over 
stationary modes of single quarks and antiquarks { (such as the 
sums (\ref{sgs}) and (\ref{sgs2}), evaluated in the next section)}, or, 
equivalently, to the bound-state propagators of these dressed model 
quarks. The sum over quark modes should naturally run up only to some 
typical hadron-physics low-energy cut-off $\Lambda \sim 0.6 - 1$ GeV. 
This cut-off on quark energies is dictated by
the fact that non-perturbative interactions between quarks operate
at low energies, whereas they gradually weaken and go over to the
perturbative regime at higher energies. 
({In the aforementioned study of ${s\bar s}$ effects of
kaon loops \cite{GeI97}, Geiger and Isgur have shown the importance 
of high-mass intermediate states in these loops.  However, since 
these are hadronic, meson--baryon intermediate states, this does not 
contradict with cut-off such as $\Lambda \sim 1$ GeV on {\it quark}
energies. Namely, the dominant portions of the results of Ref. 
\cite{GeI97} are accounted for by states lying below 3--3.5 GeV. 
For comparison, our cut-off of 1.1 GeV (see Table 1) imposed 
on the energies of one strange {\em quark} and one {\em antiquark} 
fluctuating on top of the valence nucleon state, corresponds to
total energies up to $2 \Lambda + M_N\sim$ 3 GeV as well.
This leads us to believe that we have accounted for the majority 
of important degrees of freedom.)
The cut-off values such as ours,} are 
typical of calculations in models of low-energy QCD, {\it e.g.},
the NJL model \cite{StW94}.
Obviously, we suppose
here that the nucleon strangeness is the effect of low-energy
non-perturbative QCD. Indeed, this brings us
to the question what to use
concretely for ${\cal L}_{I}$ in Eq. (\ref{formula}) in the 
explicit calculation
of $\langle N|:\overline {s}\Gamma s:|N\rangle$. 

The Lagrangian
${\cal L}_{I}$ can, of course, be any interaction that can produce
fluctuating $s \overline{s}$ pairs, but
the question is which interactions can be important in producing the
strangeness of the nucleon. For example, perturbative QCD
interactions probed in high-energy deep inelastic scattering and
revealing the sea of
$q \overline{q}$ pairs, including $s \overline{s}$,
should be relatively unimportant in this regard \cite{IoK90,DeNW93}. 
A theoretical analysis
\cite{JiT95} of the CCFR data \cite{Ba95} on strange quark
distribution functions from neutrino-nucleon deep inelastic scattering
seems to support this point of view. For example, it finds a
very small upper bound on the strange radius of the nucleon
$(|\langle r^2 \rangle_s| \leq 0.005 $ fm$^2$) \cite{JiT95}.
The  possibly enhanced nucleon strangeness
is thus expected (see, {\it e.g.} \cite{IoK90})
as an effect of non-perturbative QCD, which, at low energies, around the
nucleon mass scale, is certainly more important for hadronic structure than
perturbative QCD, and can lead to $s\bar s$ pairs already at small
momentum transfers, {\it i.e.} large distances.
Non-perturbative QCD is after all responsible for
precisely such effects as forming of a quark-antiquark condensate $\langle 0|
\bar{q} q | 0\rangle$ $(q = u, d, s)$ and a gluon condensate characterizing
the non-perturbative QCD vacuum. Some investigators (see, {\it e.g.}
\cite{GeI80}, \cite{DiP84,DiP86}, or, for comprehensive reviews,
\cite{ShBook,Sh95,ScS98}) have suggested that instantons are
among the most important non-perturbative configurations of the gluon fields.
By now it has been certainly well-established that the effective
interaction between quarks resulting from the presence of instantons (let us
call this interaction ${\cal L}_{\rm inst}$) plays a very important role in the
formation of hadron structure \cite{ShBook,Sh95}, although it is not responsible
for confinement \cite{Gr85,Si89}, as thought previously. (In the present 
approach, confinement must anyway be taken care of by the unperturbed
Hamiltonian $H_0$.) In our opinion, this ${\cal L}_{\rm inst}$ is therefore 
worth testing as an important candidate for the
interactions ${\cal L}_I$ generating the strange nucleon matrix
elements of some operators. 
A calculation \cite{StW94} in the context of the NJL model 
seems to be an indication that ${\cal L}_{\rm inst}$ is indeed
the most important part of ${\cal L}_I$.
The calculation in \cite{StW94} found that large strange-pair components 
were present in the nucleon 
only if the instanton-induced interaction was included in
low-energy dynamics. 

Here we quote the vacuum-averaged version of the instanton-induced
interaction ${\cal L}_{\rm inst}$ derived by \cite{NoVZ89} in the 
instanton-liquid approach but transformed to the $x$-space. 
It is convenient
to separate it into one-, two- and three-body pieces (Eq. (\ref{Linst}))
${\cal L}_{1}, {\cal L}_{2}$ and ${\cal L}_{3}$, respectively:

\begin{eqnarray}
{\cal L}_{1} & = & -n\left( \frac{4\pi^{2}}{3}\rho^{3}\right)
\bigg\{{\cal F}_{u}\, \bar{u}_{R}\, u_{L}
+ (u\leftrightarrow d)
+ (u\leftrightarrow s) \bigg\} \nonumber \\
& + & (R \leftrightarrow L) \; ,
\label{eq:L1}
\end{eqnarray}

\begin{eqnarray}
\lefteqn{{\cal L}_{2} =  -n\left( \frac{4\pi^{2}}{3}\rho^{3}\right)^{2}
\bigg\{ {\cal F}_{u}\, {\cal F}_{d}\,
\Big[(\bar{u}_{R}u_{L})(\bar{d}_{R}d_{L}) } \nonumber \\
 & + & {3\over 32}
(\bar{u}_{R}\lambda ^{a}u_{L}\bar{d}_{R}\lambda ^{a}d_{L}- 
\frac{3}{4} \bar{u}_{R}\sigma _{\mu \nu }\lambda ^{a}u_{L}
\bar{d}_{L}\sigma ^{\mu \nu }\lambda ^{a}d_{L})\Big] \nonumber \\
& + & (u\leftrightarrow s)
+ (d\leftrightarrow s) \bigg\}
+ (R\leftrightarrow L) \; ,
\label{eq:L2}
\end{eqnarray}

\begin{eqnarray}
\lefteqn{{\cal L}_{3} =  -n\left( \frac{4\pi^{2}}{3}\rho^{3}\right)^{3}
\, {\cal F}_{u}\, {\cal F}_{d}\, {\cal F}_{s}\, {1\over 3!}\, {1\over
{N_{c}(N^{
2}_{c} - 1)}} } \nonumber \\
& \times & \epsilon_{f_{1} f_{2} f_3} \, \epsilon _{g_{1} g_2
g_3}\bigg\{\frac{2 N_{c}+1}{2 N_{c} + 4} 
\, (\bar{q}^{f_{1}}_{R}q^{g_{1}}_{L})
(\bar{q}^{f_{2}}_{R}q^{g_{2}}_{L})(\bar{q}^{f_{3}}_{R}q^{g_{3}}_{L})
\nonumber \\ 
& + & {3\over 8(N_{c}+2)} (\bar{q}^{f_{1}}_{R}q^{g_{1}}_{L})
(\bar{q}^{f_{2}}_{R}\sigma_{\mu \nu }q^{g_{2}}_{L})
(\bar{q}^{f_{3}}_{R}\sigma^{\mu \nu }q^{g_{3}}_{L}) \bigg\}.
\nonumber \\ &&
\label{eq:L3}
\end{eqnarray}

\noindent Here, $n$ is the instanton density and ${\cal F}_{f}$'s
are the characteristic
factors (corresponding to inverse effective quark masses) composed of
current light-quark masses $m_{f}$ $(f=u,d,s)$,
average instanton size $\rho \simeq 1/3$ fm 
\cite{Sh82,DiP84,DiP86}, and the quark condensate
$\langle 0|\overline{q} q|0\rangle$.
For example, for the $u$ flavour, ${\cal F}_{u} \equiv
[m_{u}\rho - ({2\pi^{2}}/{3}) \rho^{3} \langle 0| \overline{u}
u|0\rangle ]^{-1}$, and analogously for the other flavours.
The left (and right) projected components are
defined in the usual way,
$u_{L,R} = \gamma_{\mp} u \equiv ({1}/{2}) (1 \mp \gamma_{5}) u$.

In the three-body interaction ${\cal L}_{3}$, the indices $f_i$, $g_i$
$(i=1,2,3)$ run over light flavours $u,d$ and $s$.
For example, $g_{3}=d$ means $q^{g_3}_{L} = d_{L}$.
Repeated indices are summed over. The interaction defined here by
${\cal L}_{1}, {\cal L}_{2}$ and ${\cal L}_{3}$ is actually the
same as the well-known  one of Shifman, Vainshtein and Zakharov
(SVZ) \cite{ShVZ80},
although the present three-body term (\ref{eq:L3}) looks much simpler.
In fact, Nowak \cite{No91} simply Fierzed away very complicated colour
structures present in the SVZ interaction \cite{ShVZ80}, reshuffling them
to simple prefactors involving the number of quark colours $N_{c}$.

{Although Nowak {\emph{et al.}} derived this interaction 
in the random instanton liquid model (RILM) with the help of the 
mean-field, or quenched, approximation (where the collective 
coordinates of instantons and anti-instantons are randomly 
distributed, thus neglecting potentially important correlations) 
and in the long-wavelength limit, their
version of the interaction induced by small instantons is still
considered useful even in the most recent reviews of instanton physics
\cite{ScS98}. Ref. \cite{NoVZ89} took into account the delocalization
of zero modes and long-wavelength properties (scales $>$ 1/3 fm),
arriving at the interaction basically corresponding to that of
SVZ \cite{ShVZ80}, apart from the
effects of smearing over the average size $\rho$ of a small instanton,
$\rho \simeq {1}/{3}$ fm \cite{Sh82,DiP84,DiP86}. In the limit of
no smearing the SVZ instanton-induced interaction is obtained, that is,
the interaction averaged over the small instanton volume is taken to be 
the local interaction (\ref{eq:L1})--(\ref{eq:L3}). In the long-wavelength
limit, it should approximate well the intermediate-range ($\sim 1/3$ fm) 
QCD effects, which are already of non-perturbative origin, but still not 
responsible for confinement appearing at still larger scales.}

We also note that the average instanton size
$\rho \simeq {1}/{3}$ fm $=(600 {\rm MeV})^{-1}$
is consistent with what we have said above about the typical
hadronic cut-off scale $\Lambda \sim 0.6 - 1$ GeV.
Namely, the effective interaction ${\cal L}_{\rm inst}$ cannot
be operative at energies which would probe distances
significantly smaller than the average size of these
extended objects, instantons, which produce ${\cal L}_{\rm inst}$.

Obviously, the two-body term is the one which, through Eqs.
(\ref{formula}) and (\ref{eq:contractions}), yields the graph in Fig.
\ref{fig1}.
The contribution to the nucleon
stran\-ge\-ness due to the three-body interaction ${\cal L}_{3}$ is
exemplified by the last loop in Fig. \ref{figb}.
Such graphs come about when contractions in (\ref{eq:contractions})
are performed with a strange  bilinear in ${\cal L}_{3}$.
{The ${\bar s}s$ bilinear in the one-body term ${\cal L}_{1}$ 
can produce the strange quark loops disconnected from the 
valence quarks.}

We should also comment on the consistency of using the
in\-stan\-ton-induced interaction ${\cal L}_{\rm inst}$ 
for ${\cal L}_{I}$ in Eq.
(\ref{formula}), even when we view Eq. (\ref{formula}) as
a purely perturbative result.

If we take the perturbative viewpoint, why is Eq. (\ref{formula})
applicable not only to parts of ${\cal L}_{I}$ which come from
perturbative interactions such as the perturbative gluon exchange, but
also to ${\cal L}_{\rm inst}$   (\ref{eq:L1})-(\ref{eq:L3})
which is of non-per\-tur\-ba\-ti\-ve origin? The point is that the
{\it origin} of ${\cal L}_{\rm inst}$ is non-per\-tur\-ba\-ti\-ve,
{\it i.e.} these effective interactions
between quarks are the consequence of non-perturbative gluon configurations
-- instantons. However, ${\cal L}_{\rm inst}$  itself contains a small
parameter,
namely the instanton density $n$, and it is so small that a perturbative
expansion in its powers is possible.
Original estimates \cite{Sh82} --- where
$n \approx 1.6 \cdot 10^{9}$ MeV$^{4}$ = 1 fm$^{4}$ 
 --- have proved to be reliable as they
 have remained essentially unchanged \cite{Sh95} also
in the more recent instanton-liquid calculations.
It is useful to define a ``dimensionless instanton density" 
$\tilde n$ by expressing it in
units of the average instanton size $\rho$, $n \equiv {\widetilde n}
\rho^{-4}$. The commonly accepted value is $\rho = 1/600$ MeV$^{-1}
\simeq 1/3$ fm \cite{DiP84,DiP86,Sh88}.
Therefore, $\widetilde{n} \simeq 12.4 \cdot 10^{-3} \simeq 1/81$,
and this dimensionless parameter indicates that the probability of 
finding an instanton is small.  $\widetilde{n}$ is obviously 
small enough to be used as the parameter of the perturbative expansion. 
We should also keep in mind that this is the instanton density
in the true, non-perturbative QCD vacuum, while in some circumstances the
appropriate $n$ can be even smaller. Notably, 
Ref. \cite{Kl94} has found that in the MIT bag model
enlarged with the instanton--induced interaction 
(\ref{eq:L1})--(\ref{eq:L3}), 
which is used in the next section 
for the first evaluations of the  nucleon strangeness
using formula (\ref{formula}), the instanton density is very
strongly depleted with respect to the true QCD vacuum. 

{
Of course, this depletion relative to the instanton density 
in the true QCD vacuum, is just the way to express the small 
{\it probability of penetration} of the instanton liquid from the 
RILM vacuum (modeling the true QCD vacuum supposedly outside the bag) 
into parts of the volume inside the otherwise perturbative bag. 
If one wants, this can be 
visualized as occasional penetration into the bag (with a small 
probability), of drops of the instanton liquid. It is important 
to note that it is not a different kind of liquid, with different 
properties, but that it must be the same liquid -- namely the one giving 
rise to the interaction (\ref{eq:L1})--(\ref{eq:L3}). For that reason,
we use in the instanton--induced interaction ${\cal L}_I$ 
(\ref{eq:L1})--(\ref{eq:L3}), as usual, the empirical value 
of the condensate $\langle 0 | {\bar q} q | 0 \rangle$.}

{The non-vanishing (albeit small) probability for penetration of the 
droplets of instanton liquid from the true non-perturbative ``instanton
vacuum" of QCD into the ``perturbative" MIT bag interior
explains why, in the case of the MIT bag, we should be concerned with the
stran\-ge\-ness coming from the one-body term ${\cal L}_{1}$.
It is true that this term does not involve any interaction with valence 
quarks and one would thus expect that it is already included in the 
vacuum contribution. Nevertheless, recall that in the MIT bag model, 
Donoghue and Nappi \cite{DoN86}
obtained the strangeness of the ``perturbative" MIT bag by subtracting 
the (negative) non-perturbative vacuum contribution, since everything is
measured with respect to the true non-perturbative QCD vacuum as the 
referent, ``zero" level. However, if there is a non-vanishing probability
for penetration of droplets of the random instanton liquid vacuum of QCD,
it means that the difference of the bag interior with respect to the true 
non-perturbative QCD vacuum was not so large. Hence, the over-subtracted 
strangeness should be put back in, and this is the reason for the 
one-body term ${\cal L}_{1}$ contributing to our $s\bar s$-pairs 
on top of the vacuum.
}

\section{Instanton-induced strangeness in the MIT bag model}
\label{sec4}

We now turn to the actual calculation of strange nucleon matrix elements
in a specific model, and with the instanton-induced interaction
$\mathcal{L}_{\rm inst}$ given by Eqs. (\ref{eq:L1}-\ref{eq:L3}). 
For definiteness, we quote the results for the proton---the neutron 
case is quite similar.
Using Eq. (\ref{formula}), we can write the 
proton-strangeness matrix element as
\begin{eqnarray}
\lefteqn{\langle N | : \bar{s} \Gamma s : | N \rangle = 
 i \int^{\infty}_{-\infty}
dt' \; \langle N_0 | \hat{T} : \int d^3x } \nonumber \\
&\times& \bar{s}( \vec{x}, t_0) \Gamma s ( \vec{x}, t_0):
  \; :\! \int d^3 y\: \mathcal{L}_{\rm inst}
( \vec{y}, t'): | N_0 \rangle \;,
\label{inststran}
\end{eqnarray}
where we have kept only the first term in the perturbation series 
over low instanton density.  We have treated each of the three
parts of $\mathcal{L}_{\rm inst}$ (\ref{Linst}) separately. The one-body
interaction $\mathcal{L}_1$ (\ref{eq:L1}) is the simplest of all. Since no
valence quarks take part in this interaction, the only relevant part
of $\mathcal{L}_1$ is 
\begin{equation}
-n \left( \frac{4\pi}{3}\rho^3\right) \mathcal{F}_s 
[\bar{s}_R( \vec{y}, t') \, s_L( \vec{y}, t')
 + \bar{s}_L( \vec{y}, t')\, s_R( \vec{y}, t')] \;.
\end{equation}

{Expanding the strange-quark fields $s$ like in Eq. (\ref{expandq})
and contracting them leads to the following contribution of the one-body
interaction $\mathcal{L}_1$ (\ref{eq:L1}) to the matrix element}
(\ref{inststran}):
\begin{eqnarray}
\lefteqn{\langle N | : \bar{s} \Gamma s : | N \rangle_{\mathcal{L}_1}
\;=\; 4 \pi^2 n \rho^3 \mathcal{F}_s \sum_{K,L} \frac{1}{\omega_K +
\omega_L}} \nonumber \\
\times && \!\!\! \left\{ \int d^3 x\: \bar{s}_K ( \vec{x}) \Gamma
s_{L}^{c} (\vec{x}) \int d^3 y\: \bar{s}_{L}^{c}
( \vec{y}) s_K ( \vec{y})\; + \; (s \leftrightarrow
s^c) \right\}. \nonumber \\
\label{sgs}
\end{eqnarray}

{We now choose the MIT bag as our concrete model for the 
nucleon\footnote{ The problem that the MIT bag model has with the 
breaking of the chiral symmetry on the bag boundary, is cured in 
various versions of the chiral bag model containing the pion fields 
outside the bag with quarks, complete with appropriate boundary 
conditions. In the models where the bag radius is as large as in 
the ordinary MIT bag \cite{Ja82}, the pion field outside is so 
weak that it does not perturb significantly the quark sector where our 
${\cal L}_I$ acts, and cannot influence the strangeness much. Therefore,
the results obtained in such a chirally invariant but more complicated 
model, should not be very different from those obtained in the simple 
MIT bag model, so that in the next section, we stick to the latter for 
concreteness and simplicity.
}.
Therefore, the wave functions $q_{K}(\vec{r})$ ($q=u,d,s$) denote 
the MIT bag model solutions{\footnote{We follow the conventions of 
Ref. \cite{CaOW86} for the MIT bag wave functions. See also our more
complete account \cite{KlKMP98}, where we give technicalities
in detail.}}, and $K$ stands for the set 
$\{n, \kappa, j_{3}\}$, 
where $n$ is the radial excitation number and 
the quantum number $\kappa$ is determined by 
the total and orbital angular momenta $j$ and $l$, respectively.
$\omega_K$ is the energy of the quark in the bag state $K$. 
With all this, the one-body contribution to the nucleon strangeness
is completely specified.}

The sum over $K=\{n, \kappa, j_3\}$ and $L=\{n', \kappa', j'_3\}$ 
goes up to the state with 
$n=1$, $\kappa=-1$ (corresponding to the cut-off of about 1.1 GeV),
encompassing four lowest-lying strange quark states displayed in
Table \ref{tab1}.
\begin{table}
\caption{Strange-quark energy levels $\omega_{n\kappa}$,
 which can be excited by the
instanton interaction.\label{tab1}}
\begin{center}
\begin{tabular}{|c|c|c|}
\hline\noalign{\smallskip}
$n$ & $\kappa$ & $\omega_{n\kappa}$ /MeV \\ 
\noalign{\smallskip}\hline\noalign{\smallskip}
0 & -1 & 514.0\\
0 & -2 & 726.7\\
0 & 1 & 797.4 \\
1 & -1 & 1104.9 \\ 
\noalign{\smallskip}\hline
\end{tabular}
\end{center}
\end{table}
The expression for the contribution of the two-body interaction 
$\mathcal{L}_2$ (\ref{eq:L2}) is obtained in the same way as Eq.
(\ref{sgs}). However, it is somewhat more complicated,
involving also valence quark wave functions. Luckily, the terms with
$\sigma^{\mu\nu}$ cancel out, leaving us with the proton matrix element
\begin{eqnarray}
\lefteqn{\langle N | : \bar{s}\Gamma s : | N \rangle_{\mathcal{L}_2}
\;=\; \frac{16}{3} \pi^4 n \rho^6 \mathcal{F}_q \mathcal{F}_s \sum_{K,L,\pm} 
\frac{1}{\omega_K + \omega_L}} \nonumber \\
&&\times\left\{\int d^3 x\: \bar{s}_K ( \vec{x})\Gamma
s^{c}_L ( \vec{x})\;\int d^3 y\: \bar{s}^{c}_L
( \vec{y})\gamma_{\pm} s_{K}( \vec{y}) 
\vphantom{\Bigg|}\right. \nonumber \\
&&\hspace*{1mm} \times \Bigg[
2 \bar{q}_{0,-1,\frac{1}{2}}( \vec{y})\,\gamma_{\pm}\,
q_{0,-1,\frac{1}{2}}( \vec{y})\;  \nonumber \\
&& \hspace*{20mm} + \;  \bar{q}_{0,-1,-\frac{1}{2}}
( \vec{y})\:\gamma_{\pm}\:q_{0,-1,-\frac{1}{2}}
( \vec{y})\Bigg]  \nonumber \\
&&+ \int d^3 x\: \bar{s}^{c}_K ( \vec{x})\Gamma
s_L ( \vec{x})\;\int d^3 y\: \bar{s}_L
( \vec{y})\gamma_{\pm} s^{c}_{K}( \vec{y}) 
\nonumber \\
&&\hspace*{1mm}  \times \Bigg[
2 \bar{q}_{0,-1,\frac{1}{2}}( \vec{y})\,\gamma_{\pm}\,
q_{0,-1,\frac{1}{2}}( \vec{y})\; \nonumber \\
&&\hspace*{20mm} \left.  + \;
\bar{q}_{0,-1,-\frac{1}{2}}
( \vec{y})\:\gamma_{\pm}\:q_{0,-1,-\frac{1}{2}}
( \vec{y})\Bigg] \right\}  \;.
\label{sgs2}
\end{eqnarray}
Here, $q_{0,-1,\pm\frac{1}{2}}(\vec{y})$ is the wave function 
for the ground state of the  valence quark in the bag, which 
we take to be the same for $u$ and $d$ quarks.

Going now to the three-body interaction $\mathcal{L}_3$ (\ref{eq:L3}), 
expressions
become extremely long and complicated, so we do not write them down
here. As seen below,  it turns out that this contribution is 
much smaller than the preceding two, anyway.  

After focusing on the
scalar ($\bar{s}s$) and pse\-u\-do\-sca\-lar  ($\bar{s}\gamma_5s$)
stran\-ge\-ness as the channels
preferred by the QCD-vacuum fluctuations \cite{Zh97}, we have also
checked the vector ($\bar{s}\gamma_{\mu}s$) and the axial-vector
($\bar{s}\gamma_{\mu}\gamma_5s$) channels.

The calculation of the contribution of the two-body ${\mathcal{L}_2}$
and the three-body ${\mathcal{L}_3}$
instanton interactions is tedious and 
in the manipulation
of all these formulae we have relied heavily on \emph{Mathematica}
package \cite{Wo88} for symbolic computer calculations.

To illustrate how our calculations in the MIT bag mo\-del have been
performed and in which way such a model choice influences our results, 
we briefly sketch the calculation with the one-body part 
${\mathcal{L}_1}$ interaction.

\subsection{Scalar and pseudoscalar strangeness}

Let us first consider the \textbf{scalar} strange current density
$\bar{s}s$ inside the proton. The
expression for the matrix element can be written as 
\begin{equation}
\langle N(p')|\bar{s}s e^{-i \vec{q}\cdot
 \vec{x}}| N(p) \rangle = A_{s}(q^2)
\bar{u}_N(p') u_N(p) \;,
\label{defsc}
\end{equation}
where $q^2 = (p-p')^2$, and $u_N$'s are nucleon 
spinors. $A_{s}(q^2)$ is the scalar form factor accounting at 
$q^2 =0$ for the scalar strangeness of the proton.

Calculations inside the bag model can be performed by making
the substitution $\Gamma =1$
and inserting the appropriate quark and antiquark wave functions
in (\ref{sgs}).
By a simple calculation one can show that the surviving combination 
is the one with $\kappa = -1$, $\kappa' = 1$ and $\kappa =1$, 
$\kappa' = -1$, and (\ref{sgs}) reduces to
\begin{eqnarray}
\lefteqn{\langle N | : \bar{s}s : | N \rangle_{\mathcal{L}_1}
\;=\; 4 \pi^2 n \rho^3 \mathcal{F}_s \sum_{n=0}^{1} 
\frac{4}{\omega_{n,-1}+\omega_{0,1}}}\nonumber \\
&\times&\! \bigg[ N_{-1}(x_{n,-1}) N_1(x_{0,1}) 
\int r^2 dr \nonumber \\
&\times & W_{+}(n,-1) W_{-}(0,1) j_0(x_{n,-1}\frac{r}{R})
j_0(x_{0,1}\frac{r}{R}) \nonumber \\
&& + W_{-}(n,-1) W_{+}(0,1) j_1(x_{n,-1}\frac{r}{R})
j_1(x_{0,1}\frac{r}{R}) \, \bigg]^2 \;.
\end{eqnarray}
The normalizations $N_{\pm 1}(x_{n,\pm 1})$ and the 
$W_{\pm}$--factors, related to
the quark wave functions, are given in Refs. \cite{CaOW86} and \cite{KlKMP98}.

The above equation represents the contribution to the strange 
scalar form factor 
$A_{s}(q^2 = 0)$ coming from the one-body interaction. The remaining 
contributions from the ${\mathcal{L}_2}$ and ${\mathcal{L}_3}$ instanton
interactions can be calculated similarly and the results are
\begin{eqnarray}
\langle N | : \bar{s}s : | N \rangle_{\mathcal{L}_1} &=&0.035\;, \\
\langle N | : \bar{s}s : | N \rangle_{\mathcal{L}_2} &=&0.023\;, \\
\langle N | : \bar{s}s : | N \rangle_{\mathcal{L}_3} &=&
2.9\cdot 10^{-4} \;.
\end{eqnarray}
Summing them up gives
\begin{equation}
A_{s}(0)_{\mathcal{L}_{\rm inst}} = 0.058.
\label{037}
\end{equation}
The evaluation of space-integrals has been performed numerically using the
following values: the bag radius $R$= 1/197.3 MeV$^{-1}$ $\approx$1 fm,
the average instanton size $\rho$=1/600 MeV$^{-1}$ and the instanton
density $n=2.66 \cdot 10^7$ MeV$^4$, which is the depleted instanton
density in the MIT bag as found in \cite{Kl94}. 
Moreover, we have taken the
strange quark mass $m_s$=200 MeV and the valence
quark mass $m_u=m_d\equiv m_q$=8 MeV. The quark condensate
that follows from the Gell-Mann--Oakes--Renner relation
for these quark masses and the empirical meson masses is
$\langle 0| \bar{q}q|0\rangle\approx (-200 \textrm{MeV})^3$.

The \textbf{pseudoscalar} strange form factor $B_{s}$ is defined  as 
\begin{equation}
\langle N(p') | \bar{s} \gamma_5 s e^{-i \vec{q}\cdot
 \vec{x}}| N(p) \rangle = 
B_{s}(q^2) \bar{u}_N(p') \gamma_5 u_N(p) \;.
\end{equation}
For the pseudoscalar strange current $\bar{s}\gamma_5 s$,
Eq. (\ref{sgs}) gives the vanishing one-body contribution
\begin{equation}
\langle N | : \bar{s} \gamma_5 s : | N \rangle_{\mathcal{L}_{1}}
= 0 \;.
\end{equation}
Analogously, we  obtain the vanishing result for 
the other two instanton interactions, i.e. 
$\langle N | : \bar{s} \gamma_5 s : | N \rangle_{\mathcal{L}_{\rm inst}}
= 0 \;$.
We thus obtain, 
\begin{equation}
B_{s}(0)_{\mathcal{L}_{\rm inst}} = 0 \;,
\end{equation}
as the vanishing total instanton contribution to the pseudoscalar
form factor.

\subsection{Vector and axial-vector strangeness}

In Eq. (\ref{f1f2}) the \textbf{vector} strangeness has been 
displayed in terms
of the Dirac ($F_{1}^{s}$) and the Pauli ($F_{2}^{s}$) form factors.
For comparison with experimental data, the (stran\-ge) Sachs form factors
$G^s_E$ (electric) and $G^s_M$ (magnetic) are widely used:
\begin{eqnarray}
G^s_E(q^2) &=& F^s_1(q^2) + \frac{q^2}{4 M_N^2} F^s_2(q^2) \;,\nonumber\\
G^s_M(q^2) &=& F^s_1(q^2) + F^s_2(q^2) \;.
\end{eqnarray}
By taking the non-relativistic nucleon spinor
\begin{equation}
u_N(p,s) = \sqrt{\frac{E+M_N}{2 E}} \left( 
\begin{array}{c}
\chi_s \\
\frac{\displaystyle{ \mbox{\boldmath$\sigma$} \cdot \vec{p}}}{
\displaystyle{E+m}} \chi_s 
\end{array}
\right)
\end{equation}
in the Breit frame defined by
\begin{eqnarray}
q^{\mu} &=& (q^{0}, \vec{q} ) = (0,\vec{q}_B) \;, \nonumber\\
\vec{p} &=& \frac{ \vec{q}_{B}}{2}\; , \; 
\vec{p}' = -\frac{ \vec{q}_{B}}{2} \;,
\end{eqnarray}
the components of the vector current take the form
\begin{eqnarray}
\langle N(p',s')| V_0^s | N(p,s)\rangle &=& \frac{m}{E}
\chi^{\dagger}_{s'}\chi_s G_{E}^{s}(-\vec{q}^2_B) \;, \\
\langle N(p',s')| \vec{V}^s | N(p,s)\rangle &=& \frac{1}{2 E}
\chi_{s'}^{\dagger} i ( \mbox{\boldmath$\sigma$} \times \vec{q}_B) 
\chi_s G_{M}^{s}(-\vec{q}^2_B)  \nonumber \\ \;.
\label{defGm}
\end{eqnarray}

In order to calculate the contribution of the instanton-induced vector strange 
current inside the MIT bag, we have to identify the form factors in 
(\ref{defGm}) with the Fourier-transformed vector current within the bag:
\begin{eqnarray}
\lefteqn{\langle N(p')|: V_{\mu}^{s}:| N(p) 
\rangle_{\mathcal{L}_{\rm inst}}  } \nonumber \\
& = & \langle N(p')|: \! \int d^3 r\: e^{-i \vec{q}_B\cdot 
\vec{r} } \bar{s}(\vec{r})
\gamma_{\mu} s(\vec{r}): |N(p)\rangle_{\mathcal{L}_{\rm inst}},
\end{eqnarray}
using the static limit $q \rightarrow 0$. The check with the $V^s_0$
component of the vector current gives zero, i.e. 
$G_{E}^{s}(q^2=0)_{\rm inst}= 0$, as it should be.

A similar calculation for the space components $\vec{V}^s$ shows
a non-trivial cancellation among the contributions of quarks in the
loop with different spin orientations producing the total result
\begin{equation}
G_M^s(0)_{\mathcal{L}_{\rm inst}} = 0 \, .
\end{equation}
This implies the vanishing strange magnetic moment
\begin{equation}
\mu_{s}=F_{2}^{s}(0) = 0 \;,
\label{mus}
\end{equation}
{ which is compatible with the recent measurements at MIT/Bates 
\cite{Mu97} and even more recent ones at TJNAF (JLab) \cite{An99}.}

Eq. (\ref{baryomagnetic}) then implies that
the baryomagnetic moment is 
\begin{equation}
\mu_{p}^{(0)} = 1+\kappa_{p}+\kappa_{n} = 0.88 \;
\textrm{n. m.}
\end{equation}

The estimation of the \textbf{axial-vector} strangeness can be done along the 
same lines. The form-factor 
decomposition, assuming the $G$-parity 
symmetry of the strong interactions, has the form
\begin{eqnarray}
\lefteqn{\langle N(p')|\bar{s}\gamma_{\mu}\gamma_5 s|N(p)\rangle  }
\nonumber \\
&=& \bar{u}_N(p')
\left( \gamma_{\mu}\gamma_5 G_1^{s}(q^2) + \frac{q_{\mu}}
{2 M_N}\gamma_5 G_{2}^{s}(q^2)\right) \bar{u}_N(p) \;.
\end{eqnarray}

The instanton contribution to such a matrix element can be calculated as
\begin{eqnarray}
\lefteqn{\langle N(p')|: A_{\mu}^{s}:| N(p) 
\rangle_{\mathcal{L}_{\rm inst}} }\nonumber \\ && =
\langle N(p')|: \int d^3 r e^{-i\vec{q}_B\cdot
\vec{r}} \bar{s}(\vec{r})
\gamma_{\mu} \gamma_5 s(\vec{r}):|N(p)
\rangle_{\mathcal{L}_{\rm inst}}
\end{eqnarray}
and should be compared with the axial form factors defined in the Breit 
frame as 
\begin{equation}
\langle N(p',s')|\vec{A}^{s}| N(p,s) \rangle = G_{A}^{s}(0) 
\chi_{s'}^{\dagger} \mbox{\boldmath$\sigma$} \chi_s \;.
\end{equation}
Again, it turns out that the axial-vector strangeness induced by the instanton
interaction is vanishing,
\begin{equation}
G_{A}^{s}(0)_{\mathcal{L}_{\rm inst}} = 0 \; .
\end{equation}

\section{Discussion and conclusions}
\label{sec5}

This paper deals with strange quarks at very small momentum
transfers $Q^2$, as opposed to the high values of $Q^2$, where
such non--valence components of nucleon are undisputable,
and also treatable using more standard methods of perturbative QCD 
and parton models. 
The original MIT bag model \cite{Ch74,Ch74b,DeG75} represents a
suitable starting point in predicting the low-energy properties
of low-mass hadrons. In this model, $R_{\rm bag}$  imitates the
separations $R_{\rm confining}\sim$ 1 fm at which confinement
effects are important, corresponding to the confining scale
$\Lambda_{QCD}\simeq$ 100 to 300 MeV. Short--distance effects are
supposedly taken care of by the perturbative one--gluon exchange.

However, in order to account for the effects at intermediate distances,
{\it i.e.} at momentum scales $Q \sim \Lambda_{\chi SB}\simeq$ 0.6
 -- 1 GeV, the effective interaction 
(\ref{Linst}),(\ref{eq:L1})-(\ref{eq:L3}),
induced by the liquid of small instantons
(of the average size $\rho = 1/3$ fm) appears appropriate.
Of course, the effects of the instanton--induced interactions are not
included in Donoghue and Nappi's \cite{DoN86} naive bag-model relation
(\ref{naive}) for the scalar nucleon strangeness, and the relative
importance of this naive stran\-ge\-ness and the instanton effects
is precisely what interests us here.

An advantage of formula (\ref{formula}) is that, at least in
principle, it treats
the scalar, pseudoscalar, vector, axial, tensor or pseudotensor
nucleon stran\-ge\-ness in a unified manner --- one just has to specify
what $\Gamma$ is. Within a chosen nucleon model, the evaluation
of (\ref{formula}) would proceed in essentially the same
way for each $\Gamma$, except for technical details.

In the scalar case ($\Gamma=1$), the naive bag-model stran\-ge\-ness
(\ref{naive}) is actually rather large for standard values of parameters.
For our values, given at the end of subsection 4.1, it is
\begin{equation}
A_s^{\rm Nbag} \equiv -\bra{0}\bar{q}q\ket{0} V_{\rm bag}=4.36 \;,
\end{equation}
which is much larger than the instanton-induced contribution
(\ref{037}), and dominates the summed strangeness
\begin{equation}
A_s \equiv A_s^{\rm Nbag} + A_s(0)_{\calL_{\rm inst}} = 4.42 \;.
\end{equation}

Owing to using a somewhat smaller value of the quark condensate, Do\-no\-ghue
and Nappi \cite{DoN86} obtained 3.6 for this naive strangeness,
which is still rather large. $A_s^{\rm Nbag}$ depends very
strongly on the model size parameter $R_{\rm bag}$ since
$V_{\rm bag}=R_{\rm bag}^3 4\pi/3$. For example, $A_s^{\rm Nbag}$ would
decrease by a factor of 2 if $R_{\rm bag}=0.8$ fm, a nucleon size
which may be more acceptable, as the standard MIT bag value of
1 fm seems too large ({\it e.g.}, see \cite{BrKRW88}).
However, since the model dependence on the bag radius is similar for
other presently interesting matrix elements, the model dependence
largely cancels out when one forms ratios.
In particular, the instanton-induced contribution (\ref{037})
remains small in comparison with the naive nucleon bag strangeness:

\begin{equation}
\frac{A_s^{\rm Nbag}}{A_s(0)_{\calL_{\rm inst}}} \sim 75 \;,
\end{equation}
for reasonable variations of the radius parameter.

Obviously, the contribution due to the difference in the
condensate with respect to the true, non-perturbative QCD
vacuum, dominates the stran\-ge\-ness in the nucleon bag.
Admittedly, the instanton--induced contribution of this size
{\it would} be obtained in the calculation of Eq. (\ref{037})
\emph{if}, inside the MIT bag,  the non-depleted instanton density 
$n=1.6\cdot 10^9$ MeV$^4$ were used. However, we consider
this merely as a consistency check, and not as an
alternative description of strangeness in the MIT bag. This is because using
the instanton density appropriate to the non-perturbative
QCD vacuum containing the large quark condensate,
would imply that the non-perturbative QCD
vacuum and the quark condensate were assumed not only outside, but also
inside the bag. This would indeed enable $A_s(0)_{\calL_{\rm inst}}$
to replace $A_s^{\rm Nbag}$ in full, but would also make the
MIT bag description inconsistent \cite{Kl94}.

The scalar strangeness is special because of non-va\-nish\-ing
scalar $\bar{q}q$ condensates of the QCD vacuum, which makes
it more natural that it is larger than vector, axial or
other kinds of strangeness. This is especially clear in
our approach applied to the MIT bag model. In this model, the
scalar strangeness comes mostly from the difference of
the scalar $\bar{q}q$ condensates in the true QCD vacuum
and their absence in the perturbative vacuum inside the
cavity \cite{DoN86}, while only the relatively small remainder
in the present paper
comes from the response of the valence ground state to
the strangeness--sensitive probe.
However, such a response is all that exists in the case
of the pseudoscalar, vector, axial, {\it etc.}, nucleon
strangeness, since there are no
pseudoscalar, vector, axial, {\it etc.}, QCD-vacuum
condensates either inside or outside the cavity.
Since such responses tend to be much smaller than the
non-perturbative vacuum contributions, significant
differences in magnitude between the scalar and other
kinds of strangeness are very natural in our approach.
In fact, in the present case of the MIT bag model, we find
the vanishing first--order contribution to the
vector strangeness. The vanishing first--order
contributions are also found for the pseudoscalar
and axial strangeness of the nucleon.

Thus, our results confirm the conjecture of Ref. 
\cite{Zh97} for the case of the scalar strangeness.

This makes understandable
why the results on the ``non-scalar" strange quantities,
such as the strangeness nucleon magnetic form factor
\cite{Ja89,KlP89,KlP90,PaSW91,MuB94,HoPM97,HoP93,Le96,Le95,HaMD96,Ch96}
or the strangeness electric mean-square radius
\cite{Ja89,PaSW91,MuB94,FoNJC94,HoPM97,HoP93,GeI97,Le96,Le95,HaMD96,Ch96},
vary so much, even by the sign, from one model to
another: the ``non-scalar" strange quantities should
all be rather small, and artifacts of various
models very easily put it on either side of the zero.

{
Our results are also consistent with the most recent  
measurements of the strange 
vector form factors at low momentum transfer, $Q^2 \lsim 1$ GeV.
The experimental stran\-ge magnetic form factor of the nucleon
at $Q^2 =0.1$ (GeV/c)$^2$,
$G^{s}_{M}=0.23\pm 0.37\pm 0.15 \pm 0.19$ n.m.,
obtained at MIT/Bates \cite{Mu97} is consistent with the 
absence of strange quarks, but the error bars are large. 
However, the results and conclusions of our approach, that 
channels other than the scalar one should not be appreciably
affected by strange quarks, seems to get support especially from 
the most recent and very precise TJNAF (JLab) measurement \cite{An99} 
yielding the small strange vector form factors at $Q^2 =0.48$ 
(GeV/c)$^2$,  
$G^{s}_{E}+0.39 G^{s}_{M}$ = 0.023$\pm$0.034
$\pm$0.022$\pm$0.026 n.m.
Furthermore, HAPPEX collaboration \cite{An99} plans to improve 
the accuracy of this result by a factor of two in 1999. Nevertheless,
its small central value, consistent with zero, and small errors, 
already exclude some of 
the more generous predictions \cite{Ja89,HaMD96} for the strangeness
(but not \cite{We95,MuB94} for example).   
}

\subsection*{Acknowledgement}
\footnotesize
D. K. and I. P. thank I. Zahed for getting them
started in this problem, and for many illuminating discussions.
D. K., K. K. and I. P. acknowledge the partial support of the 
EU contract CI1*--CT91--0893 (HSMU), and 
the hospitality of the Physics Department
of the Bielefeld University.

\end{document}